\documentclass{article}
\usepackage{spconf,amsmath,graphicx,hyperref}
\usepackage{amsfonts}
\usepackage{xspace}
\usepackage{booktabs}
\usepackage{cleveref}
\usepackage{arydshln} 

\crefname{section}{sec.}{secs.}
\Crefname{section}{Sec.}{Secs.}

\crefname{figure}{fig.}{figs.}
\Crefname{figure}{Fig.}{Figs.}

\crefname{table}{tab.}{tabs.}
\Crefname{table}{Tab.}{Tabs.}

\crefname{equation}{eq.}{eqs.}
\Crefname{equation}{Eq.}{Eqs.}

\usepackage{siunitx}

\newcommand{\eg}{\emph{e.g.},\ }   
\newcommand{\ie}{\emph{i.e.},\ }   

\makeatletter
\def\ps@icasspfirst{%
  \def\@oddhead{}\def\@evenhead{}%
  \def\@oddfoot{%
    \hfil\parbox[b]{\textwidth}{\centering\scriptsize
    Copyright~\copyright~2026 IEEE.  Personal use of this material is permitted. Permission from IEEE must be obtained for all other uses, in any current or future media, including reprinting/republishing this material for advertising or promotional purposes, creating new collective works, for resale or redistribution to servers or lists, or reuse of any copyrighted component of this work in other works.}\hfil}%
  \let\@evenfoot\@oddfoot
}
\makeatother

\title{LenslessMic: Audio Encryption and Authentication via Lensless Computational Imaging}
\name{Petr Grinberg, Eric Bezzam, Paolo Prandoni, Martin Vetterli}
\address{Audiovisual Communications Laboratory, EPFL, Switzerland\\ \texttt{first.last@epfl.ch}}
\begin{document}
\ninept
\maketitle
\thispagestyle{icasspfirst} 
\begin{abstract}
With society’s increasing reliance on digital data sharing, the protection of sensitive information has become critical. Encryption serves as one of the privacy-preserving methods; however, its realization in the audio domain predominantly relies on signal processing or software methods embedded into hardware. In this paper, we introduce LenslessMic, a hybrid optical hardware-based encryption method that utilizes a lensless camera as a physical layer of security applicable to multiple types of audio. We show that LenslessMic enables (1) robust authentication of audio recordings and (2) encryption strength that can rival the search space of 256-bit digital standards, while maintaining high-quality signals and minimal loss of content information. The approach is validated with a low-cost Raspberry Pi prototype and is open-sourced together with datasets to facilitate research in the area.
\end{abstract}

\section{Introduction}\label{sec:introduction}

With the rapid growth of digital connectivity, security risks such as data leaks, content manipulation, and deepfakes have become increasingly prevalent.
Conventional digital encryption, such as DES~\cite{FIPS46-3} and AES~\cite{FIPS197}, provides strong protection against malicious actors,
but complementary approaches can further strengthen defenses against emerging threats. Several tools have been designed specifically for audio data~\cite{albahrani2022review}. However, most of them rely on digital/analog signal processing or incorporate software methods into hardware~\cite{kabir2010hardware, fu2023dynamics}, while physics-based hardware techniques remain unexplored.
In this work, we explore the novel use of analog optical hardware for audio encryption and authentication, leveraging lensless computational imaging as a physical layer of security.

Integrating image-based encryption methods into the audio domain opens new frontiers for privacy. This is enabled by converting a waveform into a visual representation, such as a spectrogram or by modulating light according to acoustic vibrations~\cite{9055375}.\footnote{A principle used as early as the 1920s by \textit{sound-on-film} technologies.} Several studies have explored retrieving the signal from motion cues in a video~\cite{bilaniuk1997optical, shang2009laser, davis2014visual}. Although such systems are intended for malicious purposes such as eavesdropping~\cite{chen2024survey},
we demonstrate that 
sound-to-light conversion can be conversely used to enable new security applications for better protection and verification.
 
Recently, lensless computational imaging have been demonstrated as a robust image encryption method,
by relying on the visual privacy of lensless camera measurements~\cite{Boominathan:22,10666814} and using programmable masks as a defense against plaintext attacks and for verification~\cite{bezzam2025encryption}.
We combine the latter approach with a neural audio codec (NAC)~\cite{kumar2023high} to create a robust hybrid (software- and hardware-based) encryption method for audio signals. Our main contributions are as follows:
\begin{enumerate}
    \item We propose a novel audio-encryption method, referred to as \textit{LenslessMic}, which encrypts audio with minimal loss of content information and with high-quality decrypted signals.
    \item We show how \textit{LenslessMic} can be utilized for robust authentication of audio recordings.
    \item We validate our method with a low-cost prototype based on a Raspberry Pi sensor.
    \item We demonstrate \textit{LenslessMic}'s generalization capabilities across multiple types of audio (speech and music) and different codec settings. 
    \item We open source code, collected datasets, and demo samples to facilitate research in the area.\footnote{\url{https://blinorot.github.io/projects/LenslessMic}} 
\end{enumerate}

\section{Methodology}\label{sec:methodology}

\begin{figure*}[!ht]
    \centering
    \includegraphics[width=0.75\linewidth]{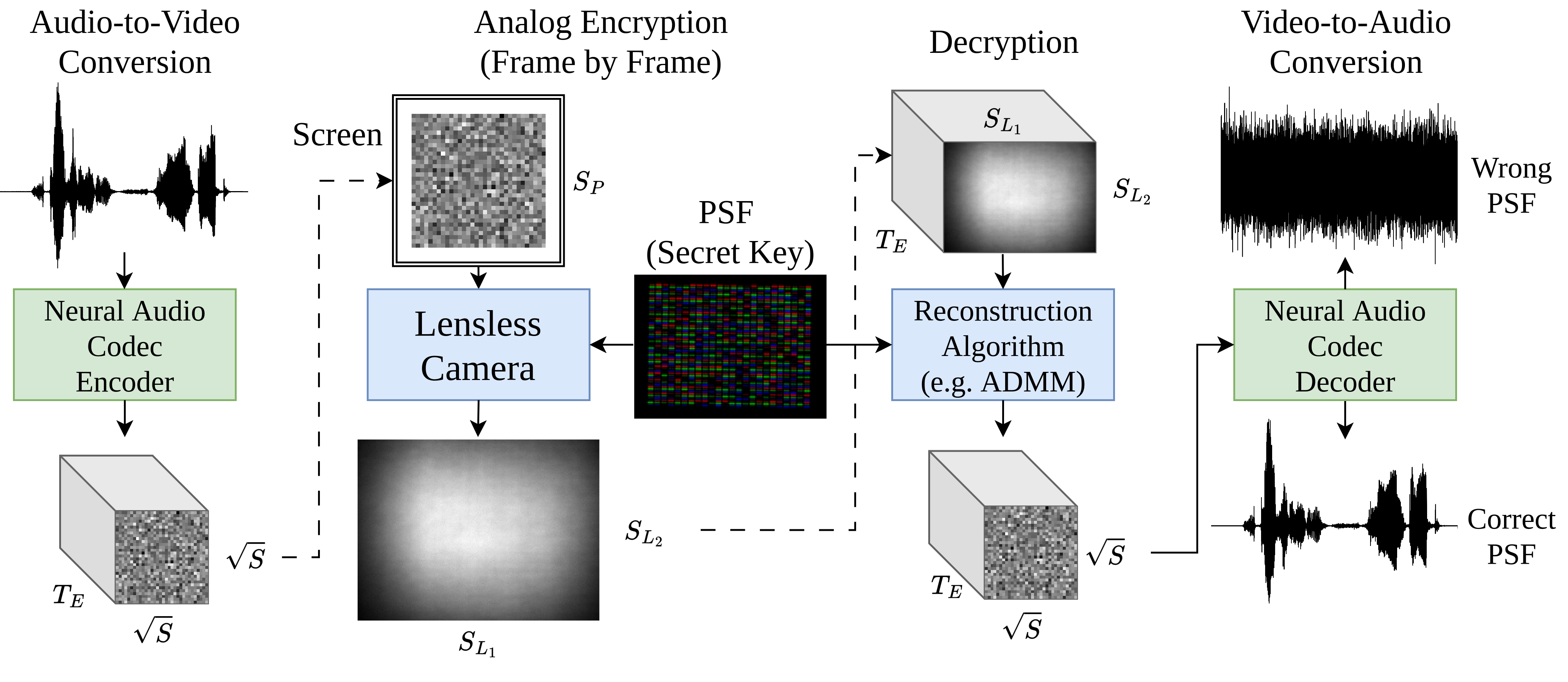}
    \caption{The proposed \textit{LenslessMic} pipeline. The waveform is converted into a time-varying visual representation that is captured frame-by-frame by a lensless camera. Reconstruction algorithm is then applied to recover audio from the lensless measurement.}
    \label{fig:lenslessmic_pipeline}
\end{figure*}

\subsection{Lensless computational imaging}

While cameras typically form images on a sensor through optics,
lensless cameras shift the image formation to the digital post-processing. The multiplexing nature of lensless cameras makes the captured measurements incomprehensible. This property is used to achieve visual privacy. The measurements can be modeled as:
\begin{equation}
\mathbf{y} = \mathbf{H}\mathbf{x} + \mathbf{n}
\end{equation}
where $\mathbf{y}, \mathbf{x}, \mathbf{n}$ are the vectorized measurement, scene intensity, and noise respectively; and $\mathbf{H}$ is the system matrix.
$\mathbf{H}$ is often assumed to have a Toeplitz structure, \ie each column corresponds to a shifted version of the on-axis point spread function (PSF),
such that the captured image can be expressed as a convolution between the scene intensity and the on-axis PSF.
With a forward model, an inverse problem can be formulated to recover a discernible image~\cite{Boominathan:22}.

Security applications of lensless cameras rely on the recovery's dependency on a correct PSF,
\ie that a perturbed or incorrect matrix $\mathbf{\hat{H}}=(\mathbf{H}-\mathbf{\Delta})$ leads to errors when decoding~\cite{bezzam2025encryption}:
\begin{equation}\label{eq:wrong_psf}
    \mathbf{\hat{x}} = \mathbf{\hat{H}}^{-1}\mathbf{y} = \mathbf{x} + \mathbf{H}^{-1}\mathbf{n} + \underbrace{(\mathbf{x}+\mathbf{H}^{-1}\mathbf{n})\sum_{k=1}^{\infty}(\mathbf{H}^{-1}\mathbf{\Delta})^k}_{\text{wrong system error}}.
\end{equation}
The larger the deviation $\mathbf{\Delta}$, the larger the image perturbation. 
Above, we demonstrate the error for direct inversion,
The same deviation can be demonstrated for regularized recovery approaches~\cite{bezzam2025towards}, \ie  Wiener filter~\cite{li2023mwdns}, FISTA~\cite{beck2009fast}, ADMM~\cite{boyd2011distributed}, and their learned variants~\cite{9239993, monakhova2019learned}.

\subsection{LenslessMic}\label{sec:method_lensless_mic}

There are several challenges that need to be addressed by an optical approach for encoding and decoding audio.
First, capturing audio with a visual microphone needs to be sensitive enough to capture the tiny and fast vibrations of acoustic signals, and might result in poorer quality due to information loss.
The sampling rate needed to faithfully capture audio is much higher than the frame rate of conventional cameras, \eg a 1-second-length signal at \SI{16}{\kilo\hertz} would be too intensive to capture optically (and reconstruct) at the sample level.
Second, with regards to encryption, 
intensity-based optical microphones~\cite{bilaniuk1997optical} are prone to information leakage,
as the signal may be recovered through 
a simple analysis of the varying brightness.

Previous work in visual microphones addresses the sampling requirements of audio by using lasers~\cite{shang2009laser} or high-speed cameras~\cite{davis2014visual}.
However, such systems can be costly (several thousand USD), thus limiting their practicality.
Our solution is to chunk audio into frames and to use a neural audio codec (NACs)~\cite{kumar2023high, zeghidour2021soundstream} to effectively compress the temporal dimension of audio, such that standard camera sensors can be used.
In our solution, the codec's latent representation is formed into an image that is projected through a lensless system for analog encryption via multiplexed measurements that can only be decoded with a ``secret key'', \ie knowledge of the system's PSF. 
An overview of our system's capture process is shown in \Cref{fig:lenslessmic_pipeline}.
The reduced temporal resolution makes it difficult for brightness-based attacks to recover a meaningful audio.
Moreover, the use of a lensless system can provide additional defense through the use of a variable PSF with a programmable mask~\cite{bezzam2025encryption}.

\subsubsection{Robust decoding}\label{sec:robust_decoding}

Computational algorithms are needed to decrypt the lensless measurement of the codec's latent embeddings, which can then be decoded for audio.
Lensless image recovery is known to exhibit significant artifacts due to inevitable model mismatch and noise amplification~\cite{bezzam2025towards}.
Fortunately, NACs can provide robustness to such noise~\cite{zeghidour2021soundstream}.
In this paper, we use the state-of-the-art DAC~\cite{kumar2023high} codec.
DAC encodes an audio signal $\mathbf{a}\in\mathbb{R}^T$ into a latent representation $\mathbf{E}\in \mathbb{R}^{T_E\times S}$, where $S$ is the latent size and $T_E$ is the number of the encoded frames. Then, DAC passes $\mathbf{E}$ through residual vector quantization (RVQ) with $C$ codebooks to get discrete codes. These codes are decoded by taking the corresponding embeddings from the codebooks and passing their sum $\mathbf{D}\in \mathbb{R}^{T_E\times S}$ through the convolutional decoder. To create the visual representation of audio, we reshape $\mathbf{E}$ into a video $\mathbf{V}\in\mathbb{R}^{T_E\times\sqrt{S}\times\sqrt{S}}$,
which is captured frame-by-frame by a lensless camera to obtain $\mathbf{L}\in\mathbb{R}^{T_E \times S_{L_1}\times S_{L_2}}$, where $S_{L_1}, S_{L_2}$ are the resolution of a camera sensor. Reconstruction techniques and extraction of the region of interest (ROI) are then applied to recover $\mathbf{\hat{V}}\simeq \mathbf{V}$ and its reshaped version $\mathbf{\hat{E}}\simeq \mathbf{E}$. To represent frames as images, we apply min-max normalization. The min-max values for each frame are provided to the reconstruction algorithm to restore the original value range.
By computing $\mathbf{V}$ from the encoder output $\mathbf{E}$ rather than from the decoder input $\mathbf{D}$ (sum of discrete code residuals), we gain additional robustness through RVQ: even if the reconstructed signal $\mathbf{\hat{V}}$ and its corresponding $\mathbf{\hat{E}}$ are imperfect, they can still map closely to the same discrete codes in the codebooks.
Alternatively, one could treat spectrograms as images for audio-to-audio conversion, but this approach would not benefit from the robustness and quantization provided by RVQ.

To enhance robustness during decoding, each frame $\mathbf{F_t}\in\mathbb{R}^{\sqrt{S}\times \sqrt{S}}$ of $\mathbf{V}$ is projected to $\mathbb{R}^{\sqrt{S_P}\times \sqrt{S_P}}$ when capturing the lensless video on a screen.
This resizing introduces ``super-pixels'' that are more distinguishable after being blurred by the lensless camera’s PSF, since the upsampling produces a grid-like structure in the recovered frames $\mathbf{\hat{R_t}}$. We choose $S_P$ in such a way that optics magnification leads to $\mathbf{\hat{R_t}}\in\mathbb{R}^{\sqrt{r^2S}\times\sqrt{r^2S}}$, \ie $r^2$-super-pixels.
After recovering an image from a given frame, we can therefore apply non-overlapping average pooling to aggregate the super-pixels and map them back into the original pixel resolution of $\mathbf{F_t}$.

\section{Experimental setup}\label{sec:experiments}

\subsection{Dataset collection}

Dataset collection is conducted following the \textit{LenslessMic} methodology from \Cref{sec:method_lensless_mic}
with the low-cost ($\approx$ 100 USD) \textit{DigiCam} prototype~\cite{bezzam2025towards} that can change its PSF as it uses a programmable mask. The latter adds another level of security that allows convenient reconfiguration in case of PSF leaks, which is not possible for fixed-mask systems~\cite{10666814}. Frames are captured with $c=8$ downsampling rate leading to $S_{L_1}=507, S_{L_2}=380$, and a super-pixel resize coefficient of $r=8$ with $S=1024$. 
For speech data, we take the subset of Librispeech~\cite{panayotov2015librispeech} recordings that are less than \SI{3}{\second}, \SI{3}{\second}, \SI{5}{\second}, and \SI{3}{\second} from the \textit{train-clean}, \textit{train-other}, \textit{test-clean}, and \textit{test-other} sets, respectively. 
For music data, we use 195 songs with permissive licenses from \textit{SongDescriber}~\cite{manco2023thesong} cropped to \SI{6}{\second}. The datasets' sizes are presented in \Cref{tab:datasets}. We use $100$ random masks for training and another $100$ for test sets. The typical lensless datasets~\cite{bezzam2025towards,monakhova2019learned} contain around 25K images, whereas we have significantly more frames in our sets. Therefore, even if the number of audio files is not that big, the evaluation is representative. 
Since NAC embeddings resemble random noise due to their compact representations, 
we also collect measurements of images sampled from $\mathcal{N}(0, 1)$.

\begin{table}[!t]
    \centering
    \scalebox{0.85}{
    \begin{tabular}{lc|cc}
    \toprule
    \textbf{Dataset} & \textbf{Split} & \# \textbf{Audio} & \# \textbf{Frames}  \\
    \midrule
    Random & train-random & 200 & 30000 \\
    Librispeech & train-clean & 587 & 73699\\
    Librispeech & train-other & 150 & 18561\\
    \midrule
    Librispeech & test-clean & 1089 & 185773\\
    Librispeech & test-other & 512 & 62901\\
    SongDescriber & test-music & 195 & 58500 \\
    \bottomrule
    \end{tabular}}
    \caption{Sizes of the collected datasets.}
    \label{tab:datasets}
    \vspace{-0.4cm}
\end{table}

\subsection{Reconstruction algorithms}

As a baseline, we use ADMM with 100 iterations~\cite{antipa2017diffusercam}. However, as ADMM alone is computationally heavy and does not account well for the model mismatch and noise,
we also apply a \textit{Learned} model that learns the parameters of 5 unrolled iterations of ADMM~\cite{monakhova2019learned} and DRUNet-style components~\cite{zhang2021plug} for PSF correction and the pre/post-processors~\cite{bezzam2025towards} (8.1M parameters total).
The noise-like structure of the codec frames and the grid-like structure of these frames on the screen suggest using SSIM~\cite{wang2004image} loss. $\mathcal{L}_{\text{raw, SSIM}}$ ensures that the $\sqrt{r^2S}\times\sqrt{r^2S}$ output of the reconstruction algorithm is a grid with $r^2$-super-pixels. We use $\mathbf{V}$ resized via the nearest neighbor interpolation as a target for this loss. $\mathcal{L}_{\text{SSIM}}$ with a smaller kernel is used to control the similarity of the noise-like structure between $\mathbf{\hat{V}}$ and $\mathbf{V}$. Since these losses focus on the structure and not on the actual values, we also adopt the mean squared error $\mathcal{L}_{\text{MSE}}$ between $\mathbf{V}$ and $\mathbf{\hat{V}}$. The final objective is the sum of these three losses.
We investigated adding an audio-based term to the loss, but it did not improve the performance.
\textit{Learned} is trained for $50$k steps with a constant learning rate of $10^{-4}$ and a batch consisting of randomly chosen 4 consecutive frames. \textit{R-Learned} follows the scheme of \textit{Learned} but uses the \textit{train-random} set for training instead of \textit{train-clean} (see \Cref{tab:datasets}).

\section{Results}\label{sec:results}

\begin{table*}[!ht]
    \centering
    \resizebox{0.85\linewidth}{!}{
    \begin{tabular}{l|cc|ccc|ccc|cc|c|cc}
    \toprule
    \textbf{Method} & \textbf{Test Set} & $\mathbf{g/r}$ & \textbf{PSNR} $\uparrow$ & \textbf{SSIM} $\uparrow$ & \textbf{MSE} $\downarrow$ & \textbf{ViSQOL} $\uparrow$ & \textbf{SI-SDR} $\uparrow$ & \textbf{Mel} $\downarrow$& \textbf{STOI} $\uparrow$ & \textbf{WER} $\downarrow$ & \textbf{SMA} $\uparrow$ & \textbf{QM-1/2} $\uparrow$\\
    \midrule
    Ground-truth & test-clean & - & - & - & - & $4.66$ & - & $0.96$ & $0.97$ & $3.03$ & $100$ & - \\
    Ground-truth & test-music & - & - & - & - & $4.51$ & - & $1.18$ & - & - & - & -\\
    Ground-truth & test-xcodec & - & - & - & - & $4.26$ & - & $1.43$ & $0.91$ & $3.22$ & $100$ & - \\
    \midrule
    ADMM & test-clean & $1/8$ & $10.82$ & $0.04$ & $25.01$ & $1.01$ & $-44.60$ & $7.95$ & $0.38$ & $100$ & $0.00$ & $0.01/0.00$ \\
    NoPSF & test-clean & $1/8$ & $17.12$ & $0.47$ & $5.79$ & $2.36$ & $-11.61$ & $2.71$ & $0.67$ & $54.25$ & $0.28$ & $4.15/0.12$ \\
    R-Learned & test-clean & $1/8$ & $19.49$ & $0.70$ & $3.41$ & $4.13$ & $7.52$ & $2.06$ & $0.95$ & $3.36$ & $\mathbf{100}$ & $28.81/5.48$ \\
    Learned & test-clean & $1/8$ & $\mathbf{22.20}$ & $\mathbf{0.85}$ & $\mathbf{1.83}$ & $\mathbf{4.50}$ & $\mathbf{9.06}$ & $\mathbf{1.14}$ & $\mathbf{0.96}$ & $\mathbf{3.31}$ & $\mathbf{100}$ & $\mathbf{39.19}/\mathbf{15.35}$ \\
    \midrule
    Learned & \begin{tabular}{c}
        test-clean\\
        (fixed min/max)
    \end{tabular} & $1/8$ & $22.20$ & $0.85$ & $2.12$ & $4.40$ & $8.53$ & $1.39$ & $0.96$ & $3.22$ & $100$ & $37.22/12.60$ \\
    \midrule
    \multicolumn{2}{c}{\textit{\textbf{Generalizing to other content/codecs}} }
    &&&&&&&&&& \\
    \midrule
    Learned & test-music & $1/8$ & $19.26$ & $0.70$ & $3.80$ & $4.29$ & $8.05$ & $1.94$ & - & - & - & $24.66/6.37$ \\
    R-Learned & test-music & $1/8$ & $\mathbf{19.70}$ & $\mathbf{0.72}$ & $\mathbf{3.33}$ & $\mathbf{4.33}$ & $\mathbf{9.43}$ & $\mathbf{1.79}$ & - & - & - & $\mathbf{32.92}/\mathbf{8.41}$ \\
    \midrule
    Learned & test-xcodec & $1/8$ & $15.38$ & $0.19$ & $21.61$ & $1.81$ & $-17.42$ & $6.08$ & $0.58$ & $68.09$ & $9.18$ & $2.68/0.04$ \\
    R-Learned & test-xcodec & $1/8$ & $\mathbf{16.35}$ & $\mathbf{0.37}$ & $\mathbf{18.42}$ & $\mathbf{2.83}$ & $\mathbf{-7.95}$ & $\mathbf{4.46}$ & $\mathbf{0.73}$ & $\mathbf{29.79}$ & $\mathbf{61.61}$ & $\mathbf{20.63}/\mathbf{3.73}$ \\
    \midrule
    \multicolumn{2}{c}{\textit{\textbf{Experiments with grouping frames}} }
    &&&&&&&&&& \\
    \midrule
    NoPSF & test-clean & $2/4$ & $16.39$ & $0.39$ & $6.92$ & $1.70$ & $-29.10$ & $3.14$ & $0.54$ & $100$ & $0$ & $1.93/0.02$ \\
    Learned & test-clean & $2/4$ & $21.38$ & $0.81$ & $2.19$ & $4.27$ & $5.97$ & $1.29$ & $0.93$ & $3.38$ & $100$ & $22.47/4.35$ \\
    NoPSF & test-clean & $3/3$ & $15.57$ & $0.29$ & $8.64$ & $1.20$ & $-40.63$ & $4.40$ & $0.32$ & $100$ & $0$ & $1.05/0.01$ \\
    Learned & test-clean & $3/3$ & $19.47$ & $0.69$ & $3.44$ & $3.65$ & $-1.27$ & $1.84$ & $0.82$ & $5.63$ & $90.08$ & $12.42/1.03$ \\
    NoPSF & test-clean & $4/2$ & $15.04$ & $0.22$ & $9.83$ & $1.15$ & $-43.45$ & $5.42$ & $0.20$ & $100$ & $0$ & $0.51/0.00$ \\
    Learned & test-clean & $4/2$ & $17.43$ & $0.52$ & $5.50$ & $2.63$ & $-9.83$ & $2.48$ & $0.66$ & $64.60$ & $1.70$ & $4.70/0.12$ \\
    \bottomrule
    \end{tabular}
    }
    \caption{Evaluation of reconstruction quality. Ground-truth rows show metrics for encoded-decoded audio (without lensless encryption). $g/r$ denotes the number of frames $g$ grouped within a lensless capture, and $r$ the super-pixel size. PSNR, SSIM, and MSE are image metrics computed against the images displayed on the screen, while ViSQOL, Mel, STOI, WER, and SMA are computed against the original audio; SI-SDR -- the codec audio. QM-1/2 shows the $\%$ of reconstructed frames that exactly match the RVQ codes up to the 1-st/2-nd codebooks.}
    \label{tab:reconstruction_results}
\end{table*}

We use PSNR, SSIM, and MSE to evaluate the quality of video reconstruction. For evaluating audio quality, we use ViSQOL, SI-SDR, and Mel distance. We also calculate STOI and WER to assess intelligibility and speaker matching accuracy (SMA) to validate if speaker identity is preserved. The state-of-the-art \textit{Parakeet-TDT-0.6B-v2} and \textit{TitaNet-L} models from NVIDIA NeMo~\cite{Harper_NeMo_a_toolkit} are used to obtain transcription and speaker embeddings, respectively. \Cref{tab:reconstruction_results} shows the metrics for \textit{ADMM}, \textit{Learned}, and \textit{R-Learned} on the test sets. We observe that \textit{ADMM} is not capable of proper frame recovery.
However, adding learnable pre- and post-processors significantly improves the algorithm and enables downstream audio recovery. While \textit{R-Learned} suffers slightly from domain shift, its audio are intelligible, which shows that training can be done on this artificial data rather than collecting for various codecs/codec settings (which can be time-consuming).
Using \textit{train-clean} (domain data) further improves the quality and achieves metrics close to ground-truth DAC encoded-decoded signals. 
Cross-type evaluation shows that methods trained on speech data are capable of recovering music data, though training on musical recordings would improve the quality. The last column of \Cref{tab:reconstruction_results} shows the $\%$ of reconstructed frames that exactly match the RVQ codes up to the 1-st/2-nd codebook. The results support the hypothesis from \Cref{sec:robust_decoding} that imperfect reconstruction can benefit from using RVQ. We note that frame-wise min/max values used for normalization can be replaced with fixed train-set average in exchange for little quality degradation, as can be seen from \Cref{tab:reconstruction_results}. 
To evaluate generalizability to other codecs, we collect \textit{test-clean} with X-Codec~\cite{ye2025codec}. While \textit{Learned} seems to overfit on DAC frames, \textit{R-Learned}'s training on random data allows it to generalize better. The listening test concludes that there are some X-Codec samples with DAC-like quality while some are very noisy. This suggests that \textit{LenslessMic} can be applied on codecs it was not trained on, however, cross-codec generalization can be improved, and it is better to train on data from a specific codec.  
Reconstructed samples are available for listening on our demo page.\footnote{\url{https://blinorot.github.io/projects/LenslessMic}}.

\subsection{Encryption}

\textit{DigiCam}'s lensless encoding changes with a different PSF, thus using an incorrect PSF leads to a poor recovery (see \Cref{eq:wrong_psf}). The PSF depends on the programmable mask's pattern, which has $N$ pixels of bit-depth $b$, \ie there are $b^{N}$ possible mask patterns. Let $W\in (0, 1]$ be the minimum ratio of correctly determined PSF pixels required to achieve intelligible audio. One can show that $W \ge \frac{K\log_b2}{N}$ leads to an encryption strength with a search space equivalent to a key size $K$~\cite{bezzam2025encryption}. In our setup, $N=1296, b=8$ meaning $W=7\%$ to match the search space of AES-256. This analysis is applicable to brute-force (BFA) or ciphertext-only (COA) attacks on the system. However, the intruder may also try chosen-plaintext (CPA) or known-plaintext (KPA) attacks, which correspond to training a neural network on data collected via a compromised camera or data leak.

\begin{figure}[!t]
    \centering
    \includegraphics[width=0.95\linewidth]{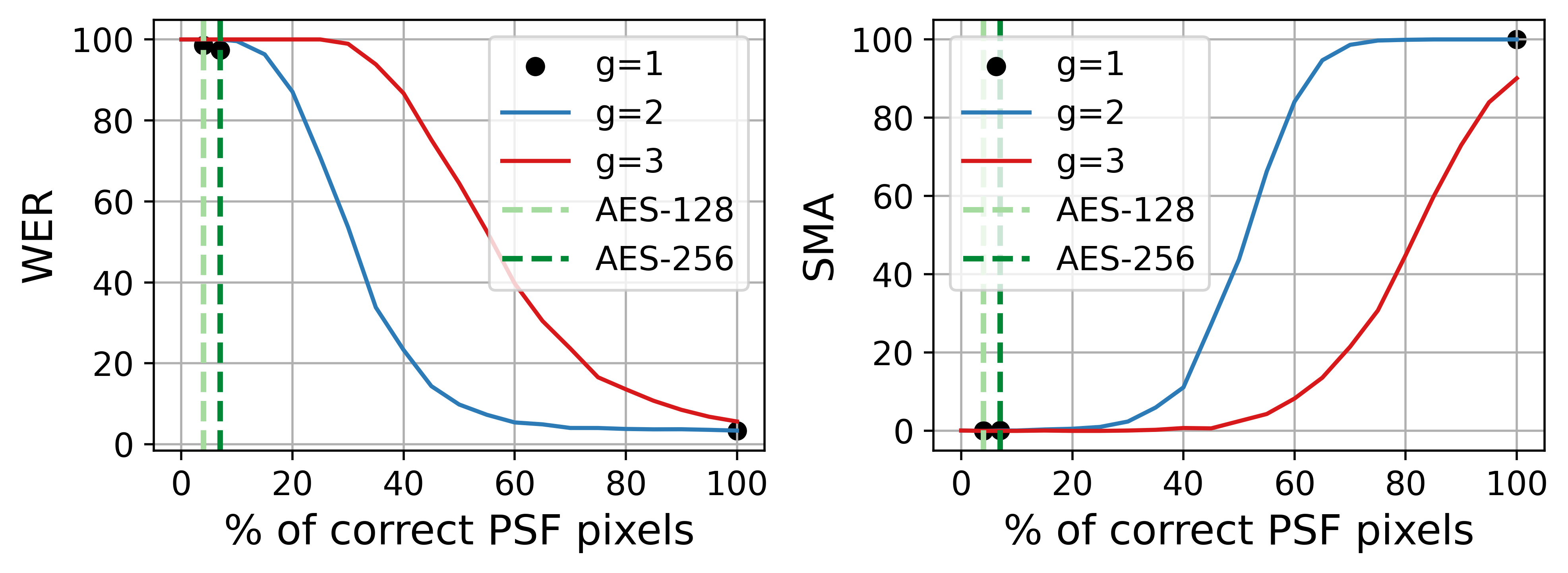}
    \caption{Reconstruction quality as a function of $\%$ correct PSF pixels.}
    \label{fig:encryption_w_results}
\end{figure}

To verify robustness against CPA and KPA, we train a DRUnet with 8.1M parameters (the same number as in \textit{Learned}) on \textit{train-clean} without provided PSF. This model is referred to as \textit{NoPSF}. However, the metrics in \Cref{tab:reconstruction_results} show that \textit{NoPSF} results in severely degraded speech, which is intelligible to some extent meaning that the content can be leaked. This is because DAC frames are only $32\times32$ in size (with large super-pixels of $r=8$), which is easier for a neural network to learn the mapping for. To address this, we combine $g^2$ consecutive frames together into $g\times g$ group, \ie $32g\times32g$ large frame and $r$ is adjusted accordingly. The system captures these large frames, which increases throughput and collection speed, while also makes the reconstruction harder. We provide metrics for $g=2,3,4$ in \Cref{tab:reconstruction_results}. Larger frames and reduced $r$ make the reconstruction much more complicated, with ($r=2, g=4$) being hard even for the \textit{Learned} algorithm. Metrics for $g=2$ suggest that this size is enough to defend against CPA and KPA since \textit{NoPSF} is not intelligible. However, manual listening of samples may allow a careful listener to understand some portions of the speech. This problem is eliminated with $g=3$, in exchange for slightly worse \textit{Learned} performance. In general, there is a trade-off between the security of the system and the quality of reconstruction.
To determine the security strength against BFA and COA, we measure how WER degrades based on $W$. \Cref{fig:encryption_w_results} shows that \textit{LenslessMic}'s search space is more robust than that of AES-256 with $W=7\%$ leading to $\text{WER}=100\%$ and $\text{SMA}=0$. Some information about the speaker and the partially intelligible text occur only after $W\ge30$. One may consider another type of BFA by comparing imperfect reconstruction with the closest RVQ output and using this output instead of the actual reconstruction. However, DAC has $C=12$ codebooks of size $1024$, \ie $2^{120}$ possible RVQ outputs. To reconstruct audio, intruder needs to repeat it for all frames, leading to $2^{120T_E}$ possible outputs, which is at least of AES-256 security level if $T_E\ge 3$. Another approach can be to fine-tune DAC with an additional 13-th codebook. Due to limited space, we provide this experiment in our demo page together with the performance analysis on \textit{test-other}.

\subsection{Authentication}\label{sec:authentication}

\begin{figure}
    \centering
    \includegraphics[width=0.63\linewidth]{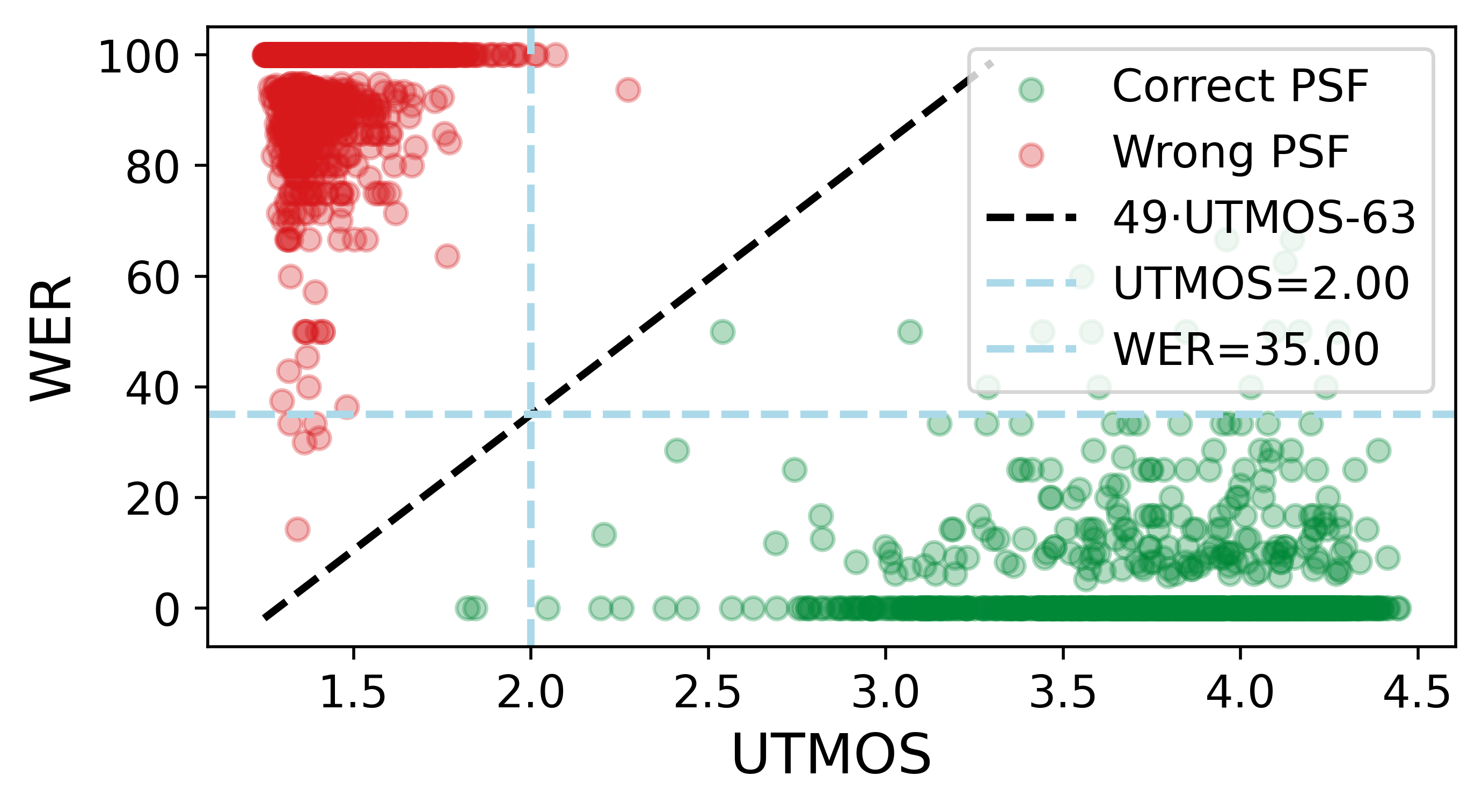}
    \caption{WER and UTMOS for correct and 10 random PSFs. Correct and wrong PSFs are perfectly separable.}
    \label{fig:authentication_results}
\end{figure}

Since a wrong PSF leads to a severely degraded signal, \textit{LenslessMic} can be used as an authentication system. Each recording is captured with a user-specific mask pattern $P_u$. To authenticate if the user $x$, who captured the recording, is the same as a test user $y$, the latter has to provide $P_y$ to the system. If $P_y=P_x$, the recovered speech should be of high quality, which is measured by some metrics, \eg WER. Metrics like UT-MOS~\cite{saeki2022utmos} can be used in case of the absence of ground-truth text/audio. We test authentication capabilities by comparing \textit{test-clean} samples with ground-truth PSF against 10 random ones. We use $35\%$ and $2.00$ thresholds for WER and UT-MOS respectively. Since random PSF results in noise, we are able to achieve $100\%$ accuracy, as can be seen in \Cref{fig:authentication_results}, when both metrics are used. Individual accuracy is $99.82\%$ and $99.95\%$ for WER and UTMOS, respectively. The results are provided for \textit{Learned} with $g=1$ but conclusion is applicable for other $g$ cases, which provide even better security. We note that $g=1$'s search space is also as secure as that of AES-256 against BFA according to \Cref{fig:encryption_w_results} meaning that it is intractable for user $y$ to guess $P_x$ to hack the system.

\section{Conclusion}\label{sec:conclusion}

In this paper, we introduced \textit{LenslessMic}, a new encryption and authentication system for audio that leverages lensless cameras to provide a physical layer of security. Experimental results demonstrate that encoded signals can be reconstructed with minimal loss while enabling robust user authentication. However, stronger security comes at the cost of reduced reconstruction quality, highlighting the need for future work on improving recovery while maintaining high security levels.
Another limitation lies in the capture process: frame acquisition is relatively slow, and lensless measurements require more storage than the corresponding audio signals. We mitigate this by packing more audio codes into each lensless capture.
Finally, the current hardware prototype, which relies on a computer monitor, is bulkier than conventional microphones or chip-based software encryption methods~\cite{kabir2010hardware, fu2023dynamics}. Future research could investigate system miniaturization, for example, through the use of compact optical components such as digital micromirror devices (DMDs).

\bibliographystyle{IEEEbib}
\bibliography{strings,refs}

\end{document}